\documentstyle[erice98,12pt,psfig]{article}
\newcommand{\be}{\begin{equation}}
\newcommand{\ee}{\end{equation}}
\newcommand{\bea}{\begin{eqnarray}}
\newcommand{\eea}{\end{eqnarray}}
\newcommand{\bit}{\begin{itemize}}
\newcommand{\eit}{\end{itemize}}
\newcommand{\non}{\nonumber}
\newcommand{\dn}{{\rm d}}
\newcommand{\re}{{\rm Re}\,}
\newcommand{\im}{{\rm Im}\,}
\newcommand{\MeV}{\:{\rm MeV}}
\newcommand{\fm}{\:{\rm fm}}

\begin{document}
\title{Two-particle properties in nuclear matter at finite temperatures}
\author{G. R\"opke and A. Schnell\footnote[1]{e-mail:
arne@darss.mpg.uni-rostock.de}}
\address{Arbeitsgruppe ``Quantenphysik und Vielteilchensysteme'',
FB Physik, Universit\"at Rostock\\
Universit\"atsplatz 1, D-18051 Rostock, Germany}
\maketitle
\abstracts{
Correlation effects in nuclear matter at finite temperatures are studied for
subnuclear densities ($\rho<\rho_0$) and medium excitation energy, where a
nonrelativistic potential approach is possible. A quantum statistical approach
is given, where clusters are treated under the influence of a clustered medium
treated within a mean field approximation. Spectral functions, in-medium cross
sections, and reaction rates are considered as well as the formation of a quantum
condensate. In particluar, exploratory calculations are shown with respect to the
pseudo gap in the nucleon level density close to the superfluid phase transition.
Estimates of isospin singlet pairing and quartetting in nuclear binding
energies are given.
}
\section{Introduction}
Nuclear matter at finite temperature is an important prerequisite to
understand the properties of not only ordinary nuclei but also
excited nucleonic systems occurring for instance in heavy-ion
collisions and astrophysical objects. Nuclear matter can be considered 
as a strongly interacting quantum liquid where the formation of
correlations is a significant feature. In particular, a simple
quasi-particle approach is not sufficient, rather one has to consider
spectral functions in order to obtain off-shell information.
We are treating equilibrium properties for
mainly symmetric nuclear matter at medium excitation energies. We will 
concentrate on two-particle correlations.

\begin{figure}[thb]
\centerline{\psfig{figure=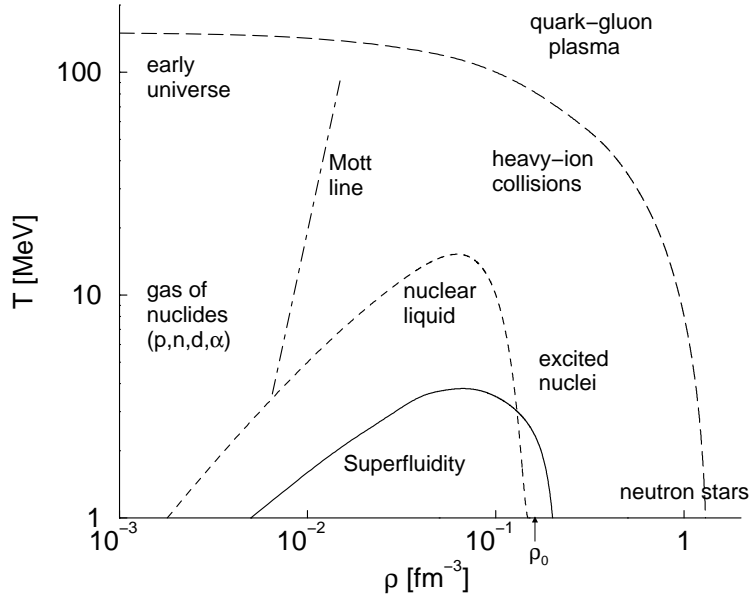,width=10cm}}
\caption{Schematic plot of the temperature-density plane of symmetric
  nuclear matter. The phase transition to the quark-gluon-plasma
  (dashed line) is hypothetical. Phase transitions strongly depend on
  isospin asymmetry.}
\label{fig-nt_plane}
\end{figure}

We are considering nuclear matter as an infinite homogeneous system
consisting of protons and neutrons. The hamiltonian $H$ is given by
\be\label{hamilton}
H = \sum_i\frac{\vec{p_i}^2}{2m_i}+\frac{1}{2}\sum_{i\neq j}V(\vec{r_i}-\vec{r_j})\,.
\ee
Such an approach is restricted to the nonrelativistic case. Different
forms are known for the interaction potential $V_{ij}$ which is
derived from nucleon-nucleon (NN) scattering data. We will use the separable
representations of the Paris and Bonn potential given by Plessas et
al.\ \cite{pestn,bestn} 
\bea
V_\alpha(p,p') & = & \sum_{i,j=1}^Nw_{\alpha i}(p)\lambda_{\alpha ij}
w_{\alpha j}(p')\,;
\quad\mbox{uncoupled}\label{V_uncoup}\,,\\
V_\alpha^{LL'}(p,p') & = & \sum_{i,j=1}^Nw_{\alpha i}^{L}(p)
\lambda_{\alpha ij}
w_{\alpha j}^{L'}(p')\,;
\quad\mbox{coupled}\label{V_coup}\,.
\eea
Due to the interaction, bound states, scattering states as well as
condensates are formed. The region of densities and temperatures where
a many nucleon system can be described by the hamiltonian
(\ref{hamilton}) is restricted to not too high densities
($\rho\le\rho_0=0.17\fm^{-1}$) and temperatures
($T\le30\MeV$) (see Fig.~\ref{fig-nt_plane}).
In this region, nuclear matter shows up
interesting phenomena such as the liquid-gas type phase transition, the
formation of bound states, and the transition to superfluidity.
\section{Cluster mean field approximation}
Different methods can be used to evaluate the properties of nuclear
matter such as path integral, computer simulations and pertubation theory using
a diagram representation for the thermodynamic Green's functions, which
will be used here.

Properties of the nucleonic many-particle system can be expressed in
terms of the thermodynamic Green's function such as the equation of
state \cite{SRS90} ($p$ denoting momentum, spin, and isospin)
\be
\rho(\beta,\mu) = \sum_p<a_p^+a_p> =
\sum_p\int\frac{\dn\omega}{\pi}f(\omega)\im G(p,\omega-i0)\,.
\ee
($f(\omega)=[\exp\{(\omega-\mu)/T\}+1]^{-1}$ is the Fermi distribution function,
$\mu$ the chemical potential, $T$ the temperature.)
The single-particle Green's function $G(p,z)$ is obtained from the
self-energy $\Sigma(p,z)$ via the Dyson equation. To include
correlations in the system, a cluster decomposition of the self-energy
is performed in terms of $n$-particle contributions
\be
\psfig{figure=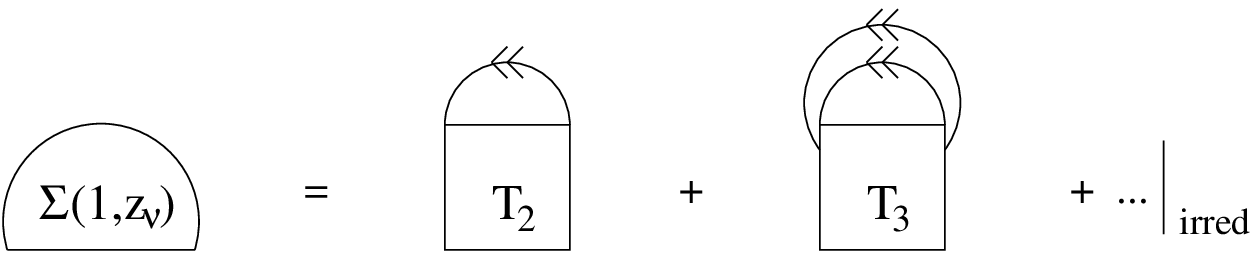,height=2cm}\;,
\ee
describing a $n$-particle cluster embedded in a mean field containing also
the contributions of correlations \cite{cluster,R94/95,DRS98}. 
In particular, we will concentrate on two-particle correlations.
The ladder approximation for the two-particle $T$ matrix is given by the
Bethe-Salpeter equation
\be\label{ladder}
\psfig{figure=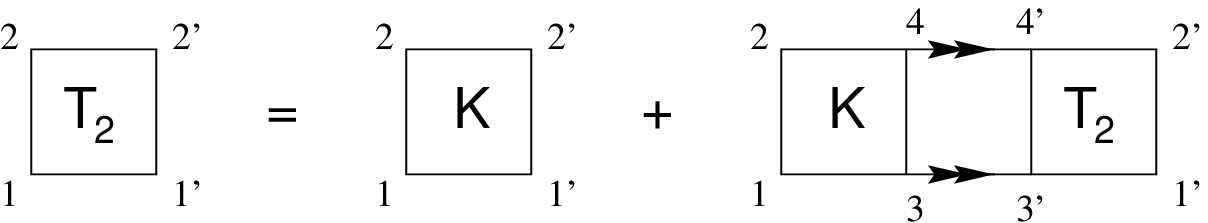,height=2cm}\;.
\ee
The interaction kernel $K$ contains, besides the potential $V$, also terms
which describe the influence of the medium in consistency with the
approximation made for the self-energy $\Sigma$ (Ward identities).
In particular, the one-particle Green's function is given in terms
of the self-energy via the Dyson equation
\be
\psfig{figure=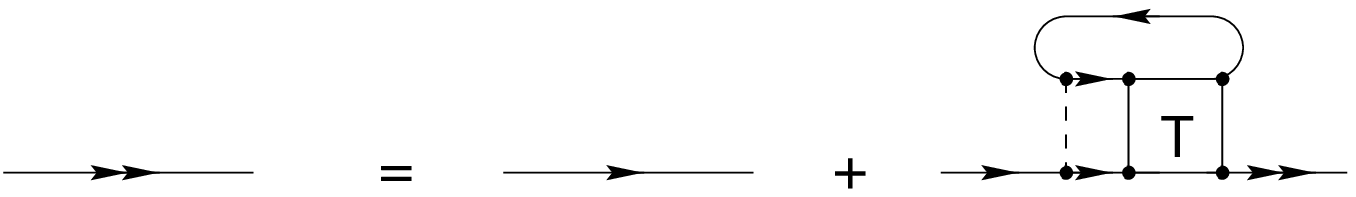,width=13cm}\;.
\ee
For the calculation of the self-energy the two-particle Green's function
has to be calculated as
\be
\psfig{figure=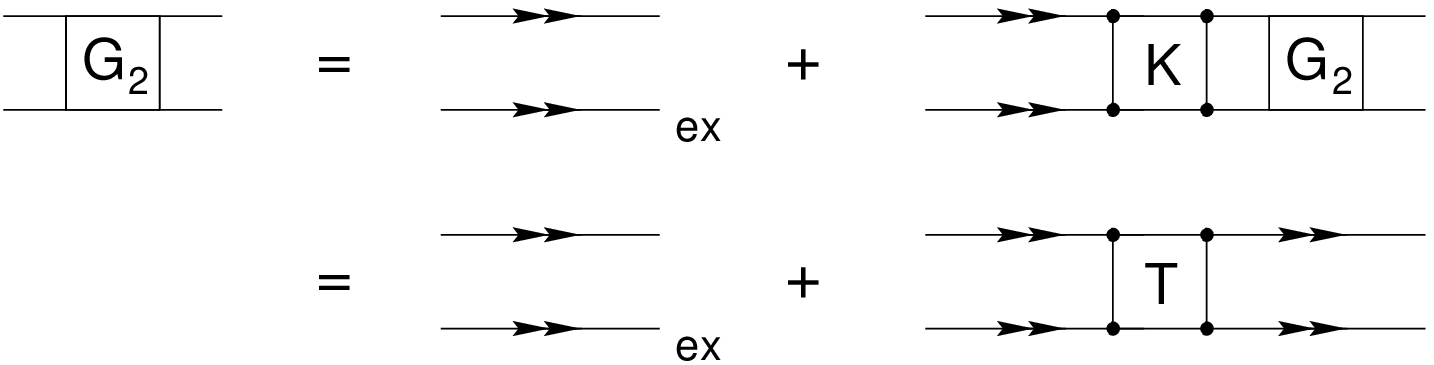,width=14cm}\;.
\ee
The interaction kernel $K$, which is related to the $T$ matrix according to
(\ref{ladder}), is consistently given by
\be
\hspace*{-1.2cm}\psfig{figure=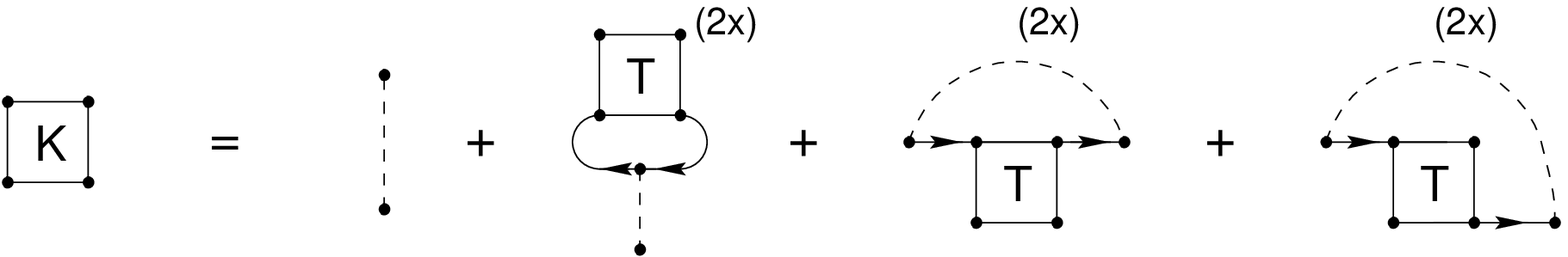,width=17cm}\;.
\ee
The ordinary Hartree-Fock type approach is extended by including in $\Sigma$
the two-particle correlations occurring in the medium
but also, in a consistent manner, in $K$. The systematic account of
$n$-particle correlations is denoted as cluster-mean field
approximation. The correlations in the medium have to be determined
in a self-consistent way \cite{cluster,R94/95,DRS98}.

In a simplified approach the equations above are not solved self-consistently,
but treating the two-particle propagator in the Bethe-Salpeter equation for
the $T$ matrix on a quasi-particle level.
Furthermore the Br\"uckner approach is reproduced neglecting the contribution in
$K$ beyond $V$.
\section{Two-particle properties in nuclear matter}\label{twopart}
\begin{figure}[hbt]
\begin{minipage}{8.75cm}
\psfig{figure=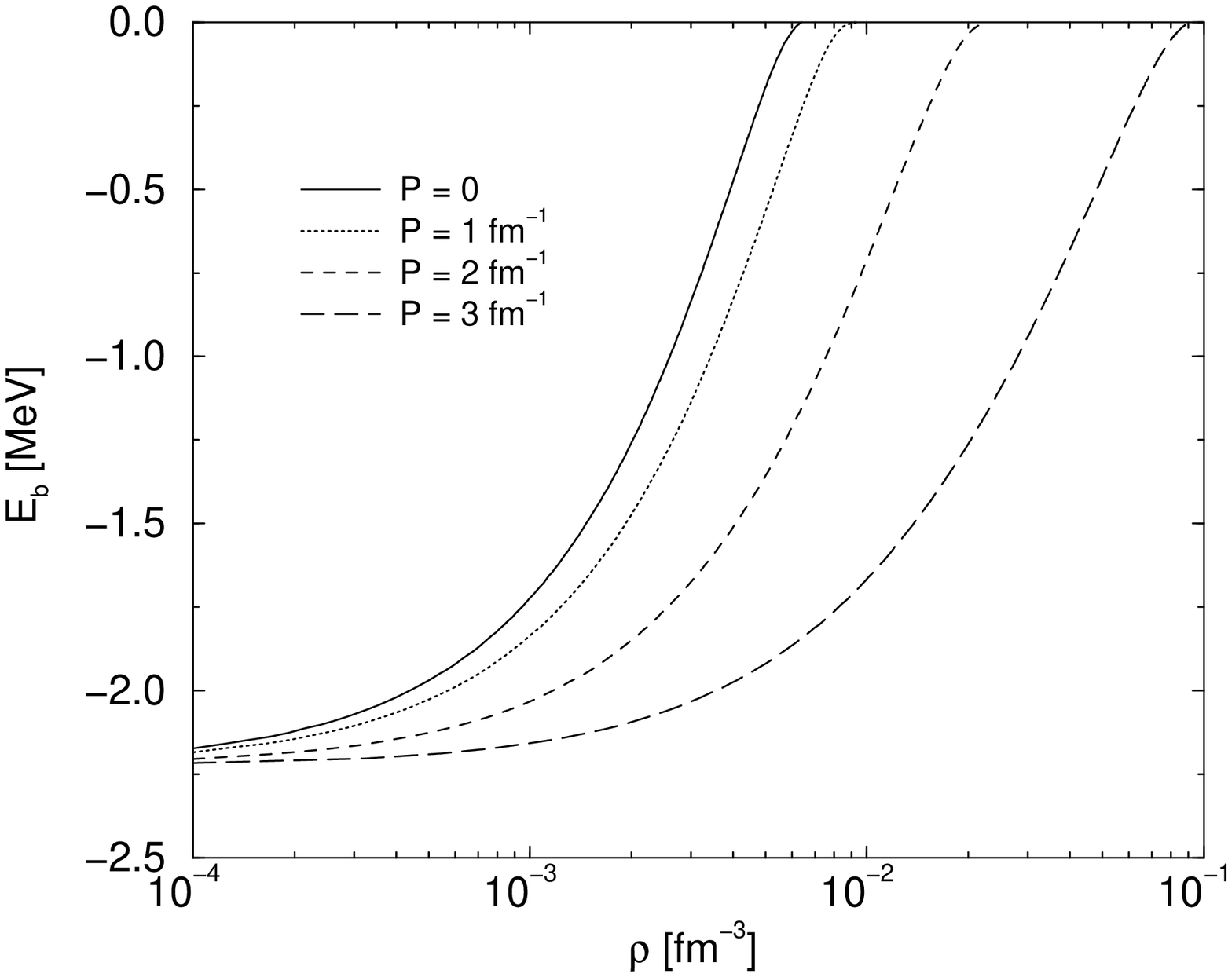,height=7.0cm}
\caption{Deuteron binding energy in the nuclear medium as a function of the
density for several total momenta. The calculation was done at $T=10\MeV$
temperature using the PEST4 potential (cf. \cite{SRS90}).}
\label{fig-eb}
\end{minipage}
\hfill
\begin{minipage}{8.75cm}
\psfig{figure=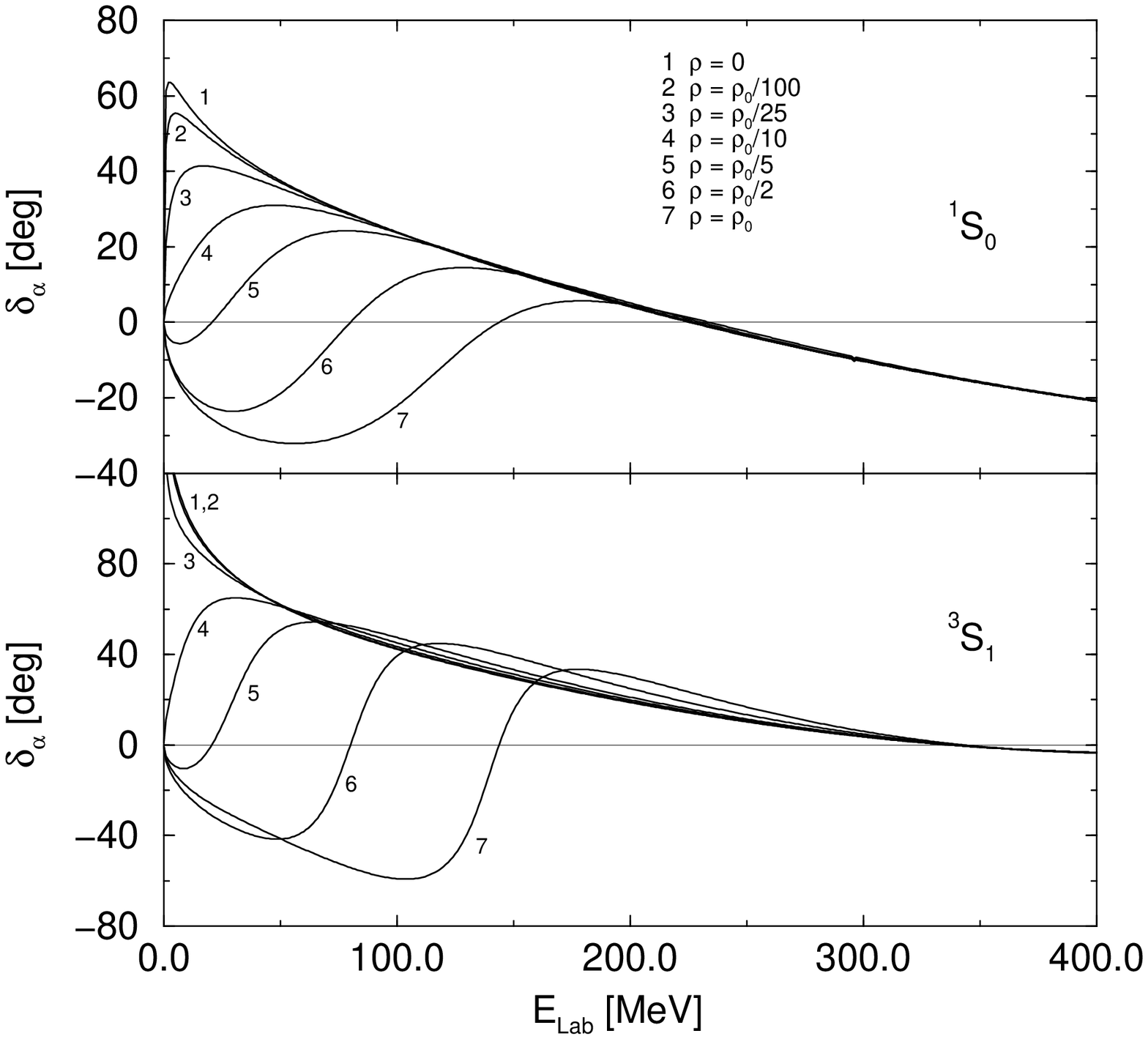,height=7.4cm}
\caption{In-medium NN scattering phase shifts as a function of the lab energy
for several values of the density at the temperature $T=10\MeV$.}
\label{fig-phase}
\end{minipage}
\end{figure}

Using the separable form of the potential (\ref{V_uncoup}),(\ref{V_coup})
an explicit
solution of the Bethe-Goldstone equation in quasi-particle approximation can
be given. Within a channel decomposition the following expressions for the $T$
matrix can be given \cite{SRS90,spec-high,spec-low,zphys}
\be
T^{LL'}_\alpha(p,p',P,z) = \sum_{ijk}
w_{\alpha i}^L(p)
[1-J^\alpha(P,z)]_{ij}^{-1}
\lambda_{\alpha jk}w^{L'}_{\alpha k}(p')\,\,, \label{Tsep}
\ee
where the quantity $J^\alpha_{ij}$ ($\alpha$ denoting the interaction channel)
is given by
\be
J^{\alpha}_{ij}(P,z) = \int\limits_0^{\infty}\dn p\,p^2\sum_{n L}
w_{\alpha n}^L(p)\lambda_{\alpha in}w_{\alpha j}^L(p)
G_2^0(p,P,z)\,,\label{J}
\ee
with the two-particle Green's function in quasi-particle approximation
\be\label{G20-qp}
G_2^0(p,P,z) = \frac{Q(p,P)}{\frac{p^2}{m}+\frac{P^2}{4m}+u(p,P)-z}\,.\label{G20}
\ee
$Q(p,P)$ and $u(p,P)$ are the angle-averaged Pauli blocking and single-particle
potential, respectively, depending on relative ($p$) and total ($P$) momentum.
From Eq.~(\ref{Tsep}) one can easily identify the poles of the $T$ matrix.
Considering negative energies one obtains the following condition for the occurrence
of a pole which describes in the $^3{\rm S}_1-^3{\rm D}_1$ channel the deuteron
bound state
\be\label{eb_cond}
\det[1-\re\,J^{\alpha}(P,\omega=E_b(P)+E_{\rm cont}(P))]_{ij} = 0\,;
\qquad(\alpha={}^3{\rm S}_1-{}^3{\rm D}_1)\,.
\ee
The solution of this equation is given in Fig.~\ref{fig-eb} where the in-medium
binding energy $E_b$ of the deuteron ($E_{\rm cont}(P)=\frac{P^2}{4m}+u(0,P)$)
is plotted for several values of the total momentum as
a function of the density. With increasing density the binding energy drops to zero
at the so-called Mott density. This means that in the nuclear medium the deuteron
is dissolved if the density is larger than approximately $\rho_0/20$. A finite
total momentum leads to a shift of this Mott effect towards higher densities.
Another quantity that can be derived from Eq.~(\ref{Tsep}) is the in-medium
NN scattering phase shift. It can be obtained by
\be
\cot{\delta_\alpha(\omega)} = \frac{\re T(ppP,\omega)}{\im T(ppP,\omega+i0)}\,.
\ee
In Fig.~\ref{fig-phase} the phase shifts of the $^1{\rm S}_0$ and
the $^3{\rm S}_1$ channel are given as a function of the energy
for various densities at $T=10\MeV$ temperature. For very low densities the phase
shifts approach the limit of free scattering. For higher densities the phase
shifts are strongly modified in the low-energy region. In particular, one
observes a jump of the $^3{\rm S}_1$ phase shift at $\omega=0$ from $\pi$ to
zero at approximately $\rho_0/20$. According to the Levinson theorem
$\delta_{\alpha}(P,\omega=0) = n\pi$ ($n$ -- number of bound states)
this is due to the vanishing of the deuteron bound state at this density value.
\section{One- and two-particle spectral function}
\begin{figure}[thb]
\centerline{\psfig{figure=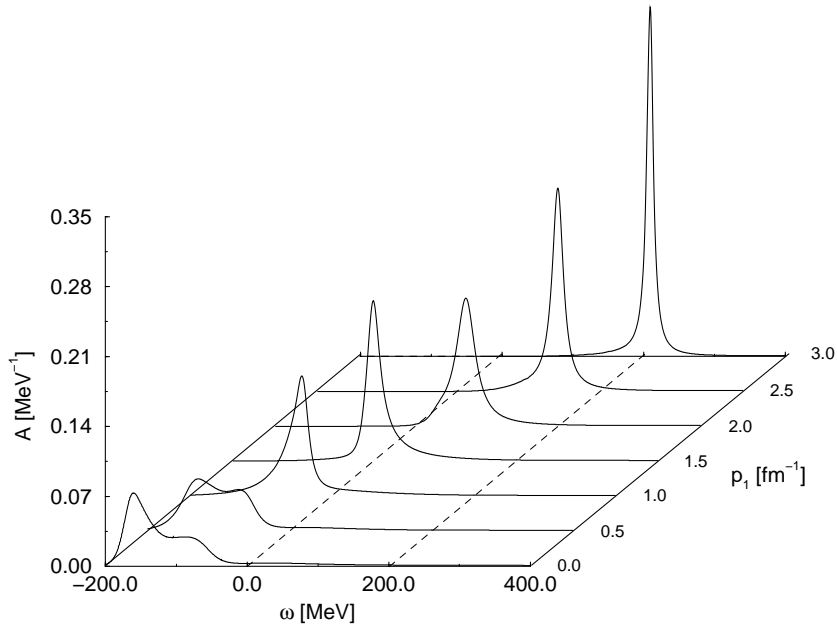,width=11cm}}
\caption{Nucleon spectral function as a function of the energy $\omega$ and
momentum $p_1$ at the density $\rho=\rho_0$ and temperature $T=10\MeV$
\cite{spec-high}. The NN potential used is a separable form of the Paris
potential \cite{pestn}.}
\label{fig-spec}
At Fermi momentum one observes a
Lorentz-like shape with a reduced but still considerable width. This
is due to the fact that at finite temperatures the quasi-particles have finite
life-time even at the Fermi surface.
\end{figure}

\begin{figure}[hbt]
\begin{minipage}{8.75cm}
\centerline{\psfig{figure=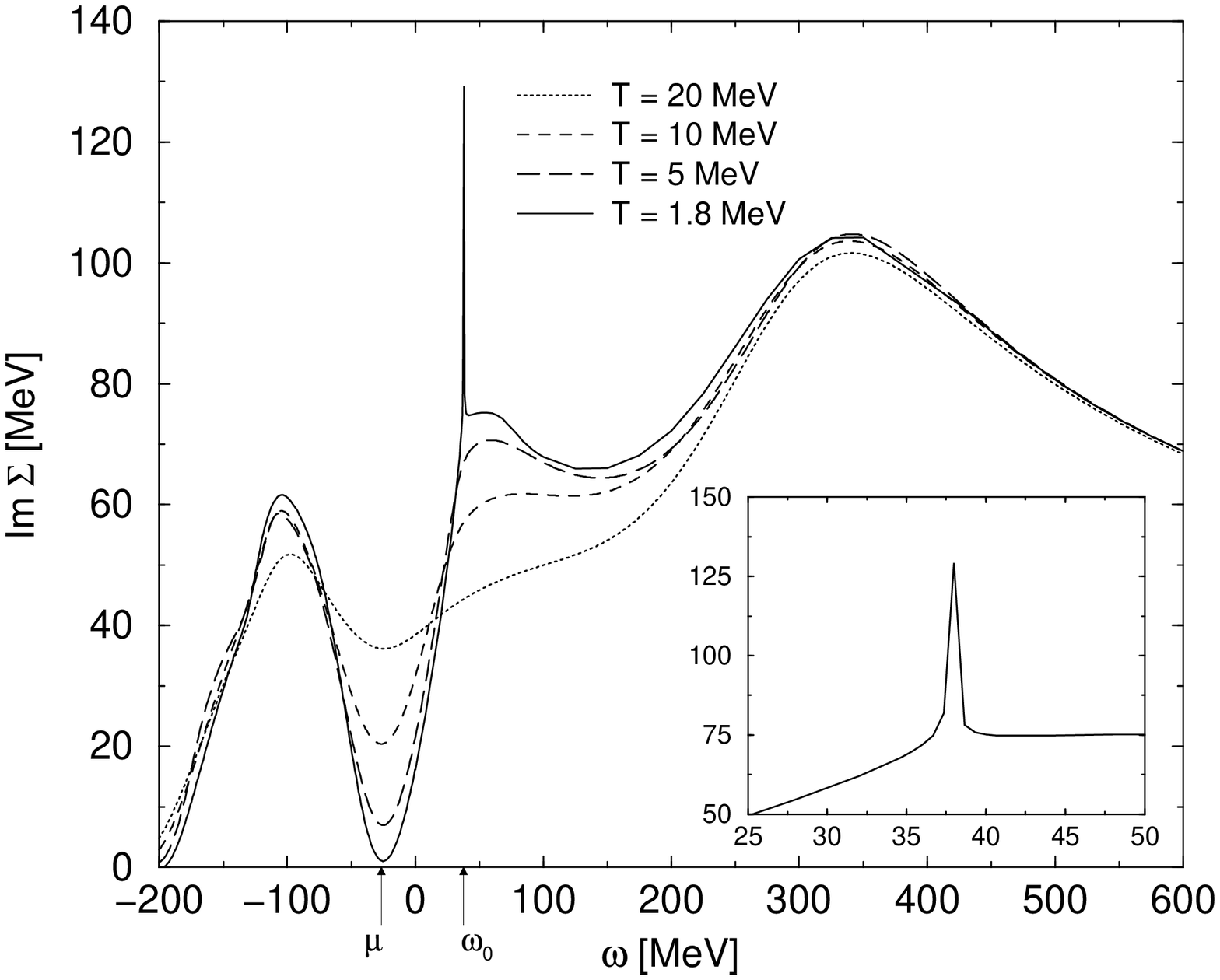,height=7.1cm}}
\caption{Imaginary part of the self-energy as a function of the energy for
several temperatures at a fixed density $\rho=\rho_0$. The momentum was set
to zero. The critical temperature for the onset of superfluidity is $T_c=1.78\MeV$.}
\label{fig-imsig}
\end{minipage}
\hfill
\begin{minipage}{8.75cm}
\centerline{\psfig{figure=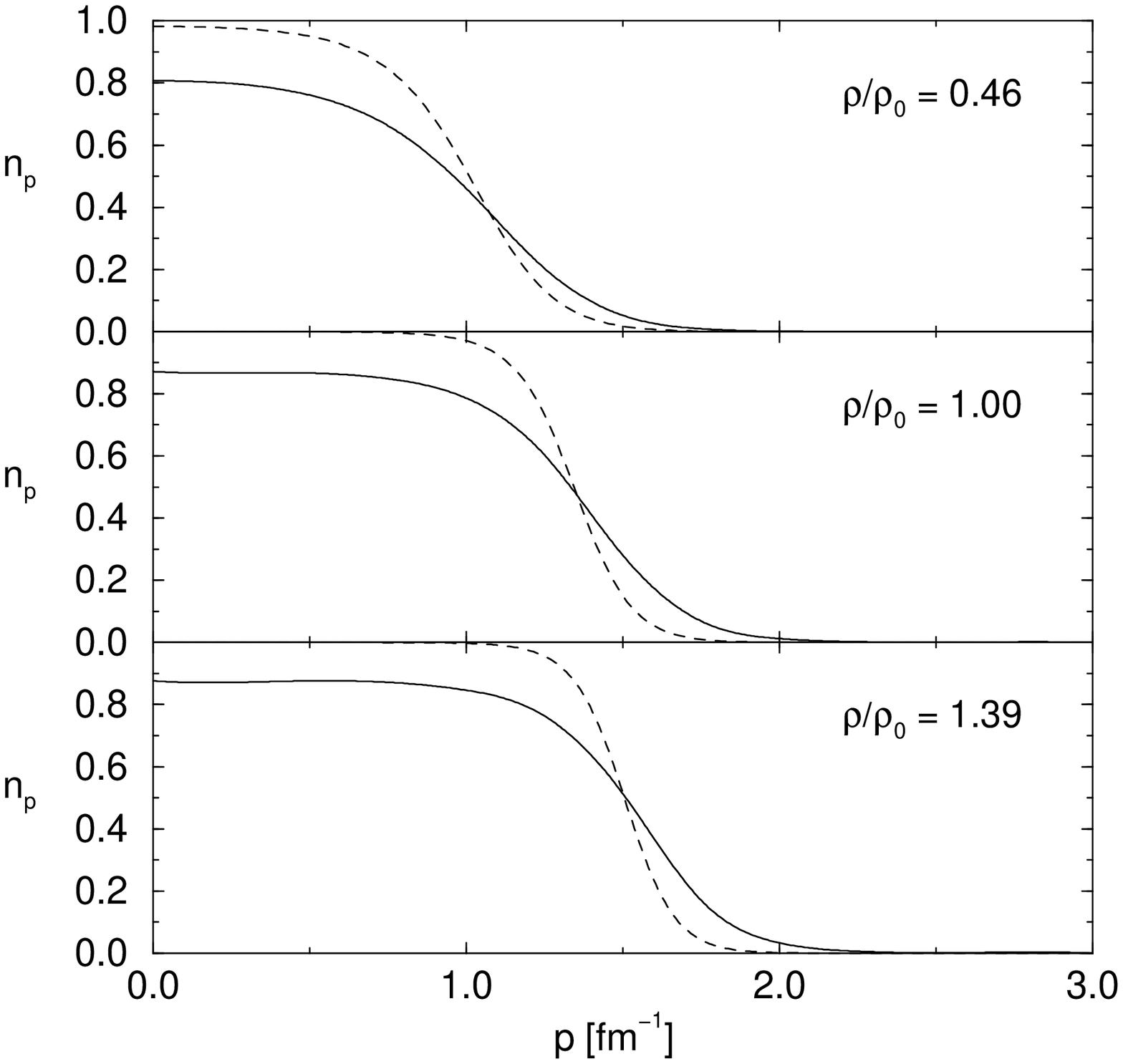,height=7.2cm}}
\caption{Nucleon momentum distribution for three values of the density for
the temperature $T=10\MeV$. The dashed curves represent the ideal Fermi gas.}
\label{fig-npn}
\end{minipage}
\end{figure}

The description of correlations in nuclear matter requires the knowledge of the
full off-shell behavior of the nucleons.
This information is contained in the spectral function of the nucleon. 
The solution of the two-particle problem can be used to improve the approximation
for the spectral function. In this sense we go beyond the quasi-particle picture.
The single-particle spectral function $A_1$ is calculated via the nucleon
self-energy which itself is given by the $T$ matrix in ladder approximation.
It reads
\be
A_1(p,\omega) = \frac{2\im\Sigma(p,\omega+i0)}{[\omega-p^2/(2m)-
\re\Sigma(p,\omega)]^2+[\im\Sigma(p,\omega+i0)]^2}\;.
\ee
In Fig.~\ref{fig-spec} the nucleon spectral function is displayed as a function
of the energy and momentum \cite{spec-high}.
The density is set to normal nuclear matter density $\rho_0=0.17\fm^{-3}$,
the temperature is $T=10\MeV$. For low momenta $A_1$ has a width of about $150\MeV$
which is quite large, particluarly if compared to the $\delta$-distribution of
the quasi-particle spectral function.
This can also be seen in
Fig.~\ref{fig-imsig} where the imaginary part of the self-energy is plotted as
a function of the energy for various temperatures. At the chemical potential
$\mu$ it exhibits a minimum which touches ground only for $T\rightarrow 0$.
Another interesting feature to be seen in Fig.~\ref{fig-imsig} is the sharp
peak at $\omega_0=2\mu-\epsilon(0)$ for the lowest temperature $T=1.8\MeV$.
It is a direct
consequence of the pairing instability in the $T$ matrix and therefore a
precursor effect of the onset of superfluidity in nuclear matter which occurs
at the critical temperature $T_c=1.78\MeV$ (see Sect.~\ref{superf}).
From the spectral function one can easily obtain the nucleon momentum
distributions via
\be
n(p) = \int\frac{\dn\omega}{2\pi}f(\omega)A(p,\omega)\,.
\ee
In Fig.~\ref{fig-npn} the momentum
distribution is given for three values of the density and compared to the
respective quasi-particle Fermi distribution functions. At low momenta one
observes a depletion by about 10 to 15 percent, whereas for high momenta the
momentum distribution is enhanced. This behavior is well known from electron
scattering experiments (e,e'p) and due to nucleon-nucleon correlations. Similar
investigations can be done for low-density nuclear matter where the deuteron
bound state arises yielding an important contribution to the spectral
function which cannot be described within a quasi-particle model \cite{spec-low}.

\begin{figure}[hbt]
\begin{minipage}[c]{8.75cm}
\psfig{figure=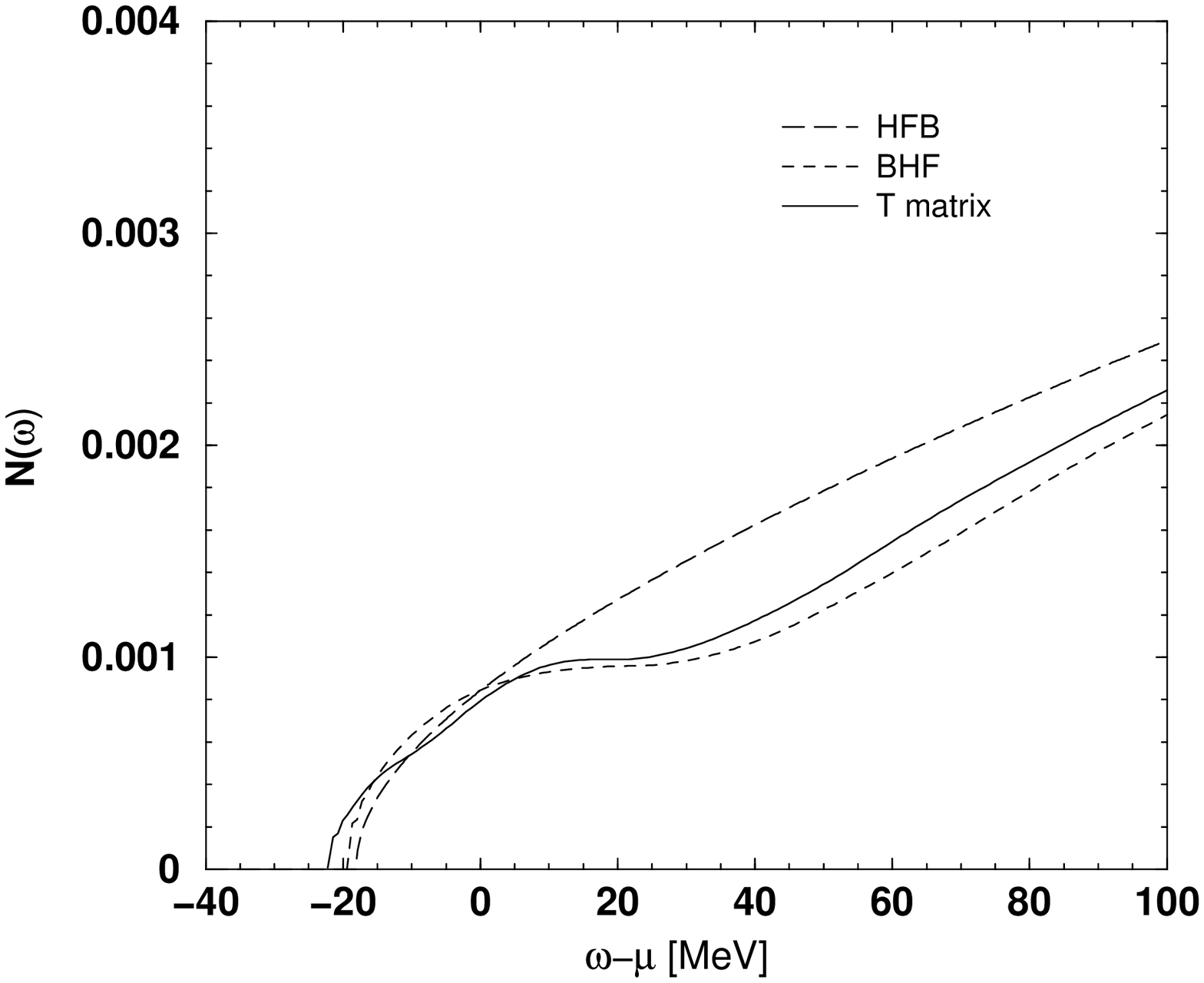,width=8.3cm}
\end{minipage}
\hfill
\begin{minipage}[c]{8.75cm}
\psfig{figure=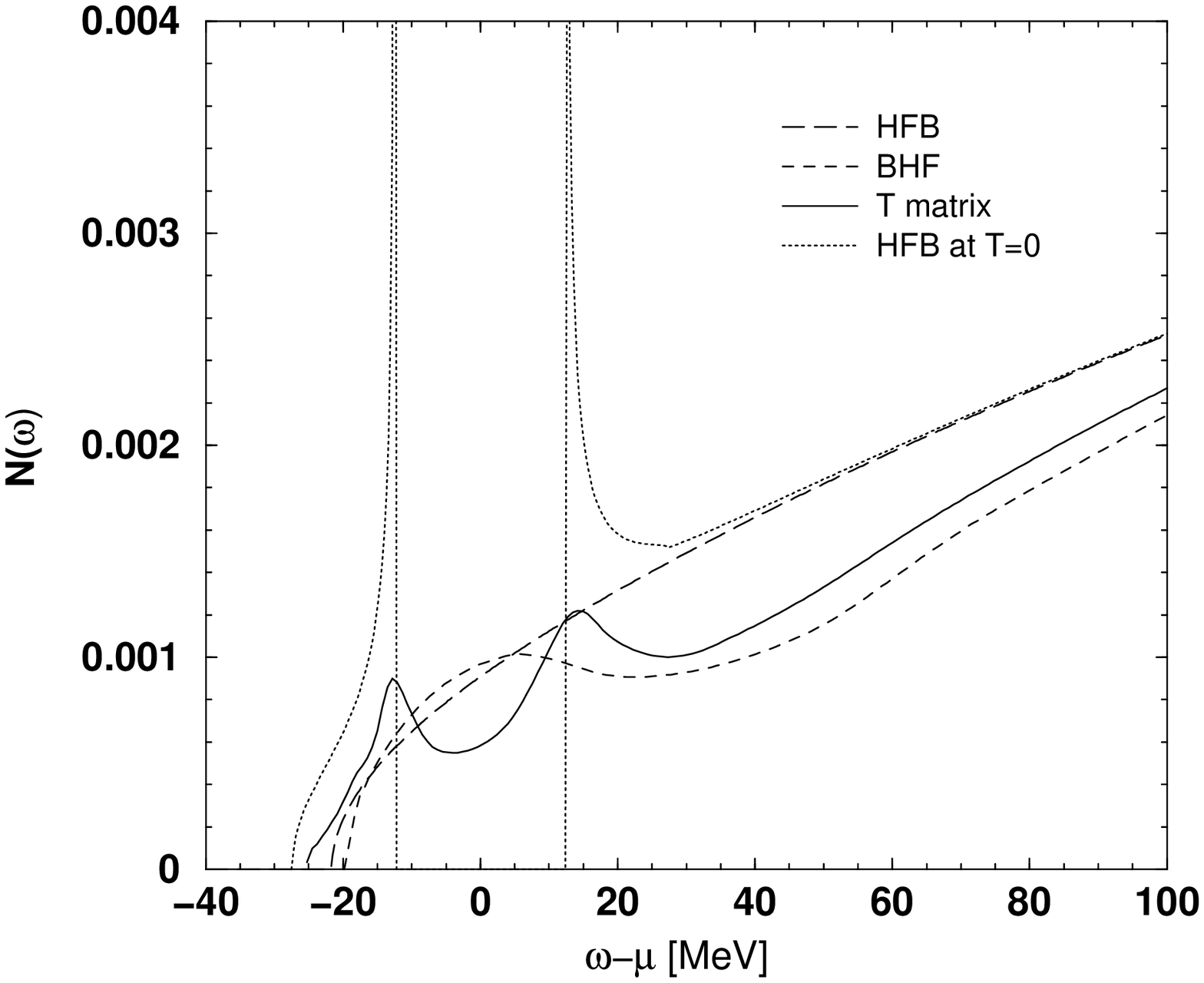,width=8.3cm}
\end{minipage}
\caption{Nucleon level density $N(\omega)$ as a function of $\omega-\mu$
($\mu$ -- chemical potential) at the density $\rho=\rho_0/3$ for different
approximations (HF-Bogoliubov, Br\"uckner-HF, $T$ matrix approach). Left:
Temperature $T=10\MeV$, far above the critical temperature; Right:
Temperature $T=6\MeV$, close to the critical temperature. For comparison
the BCS solution at $T=0$ is given by the dotted curve.}
\label{fig-leveldens}
\end{figure}

In Section \ref{twopart} it has been demonstrated that the deuteron bound state
as well as scattering states play an important role in low-density nuclear
matter. Similar to the description of single-particle states the two-particle
spectral function is the appropriate tool for two-particle states. It is
defined by the imaginary part of the two-particle Green's function
$a_2(p_1p_2p_1'p_2',\omega)=\im G_2(p_1p_2p_1'p_2',\omega)$ and can
be also calculated via the ladder $T$ matrix.
In order to have the spectral function as function of total momentum and
energy at our disposal and for practical purposes we define
\be
A_2(P,\omega) = \sum_{pp'}w(p)a_2(P/2+p,P/2-p,P/2+p',P/2-p',\omega)w(p')\,.
\ee
which is an average with respect to the relative momenta retaining
the important informations. Of particular interest is the behavior of $A_2$
close to the superfluid phase transition. In contrast to BCS theory, which can
be applied above and below $T_c$, but is based on quasi-particle concepts,
the standard $T$ matrix approach includes correlations, but cannot be
extended to describe superfluid nuclear matter. As a first step to overcome
this deficiency we investigate the precritical behavior of the level density
\be
N(\omega) =  \sum_{p}\delta(\omega-\epsilon(p))\,,
\ee
which is given by the real part of the nucleon self-energy and implicitly
by the two-particle spectral function
\be
\epsilon(p) = \frac{p^2}{2m}+U(p)\,;\qquad
U(p) = \re\Sigma(p,\epsilon(p))\,.
\ee
The real part of the self-energy can be cast in the following form
\be
\re\Sigma(p_1,\epsilon(p_1)) = \sum_2f(\epsilon_2)
\re T(p_1p_2p_1p_2,\epsilon_1+\epsilon_2)
+\sum_2\int\frac{\dn\omega}{\pi}g(\omega)
\frac{\im T(p_1p_2p_1p_2,\omega+i0)}
{\omega-\epsilon_1-\epsilon_2}\,.\label{resig}
\ee
In Fig.~\ref{fig-leveldens} the level density is given as a function of the
energy for two temperatures. At $T=6\,\MeV$ (close to the critical
temperature of the superfluid phase transition) the level density exhibits
a pre-critical pseudo gap which has the size of the BCS gap at $T=0$. This
means, that the onset of superfluidity does not happen sharply at $T_c$
signaled by the spontaneous opening of a gap in the level density as supposed
by BCS theory. Rather, due to strong correlations, a soft transition to
superfluidity with the formation of a pseudo gap can be observed. It is
therefore necessary to go beyond the mean field approach of the BCS-theory
and to generalize the $T$ matrix approximation to below $T_c$.
\section{In-medium cross sections}
\begin{figure}[hbt]
\begin{minipage}[t]{8.75cm}
\psfig{figure=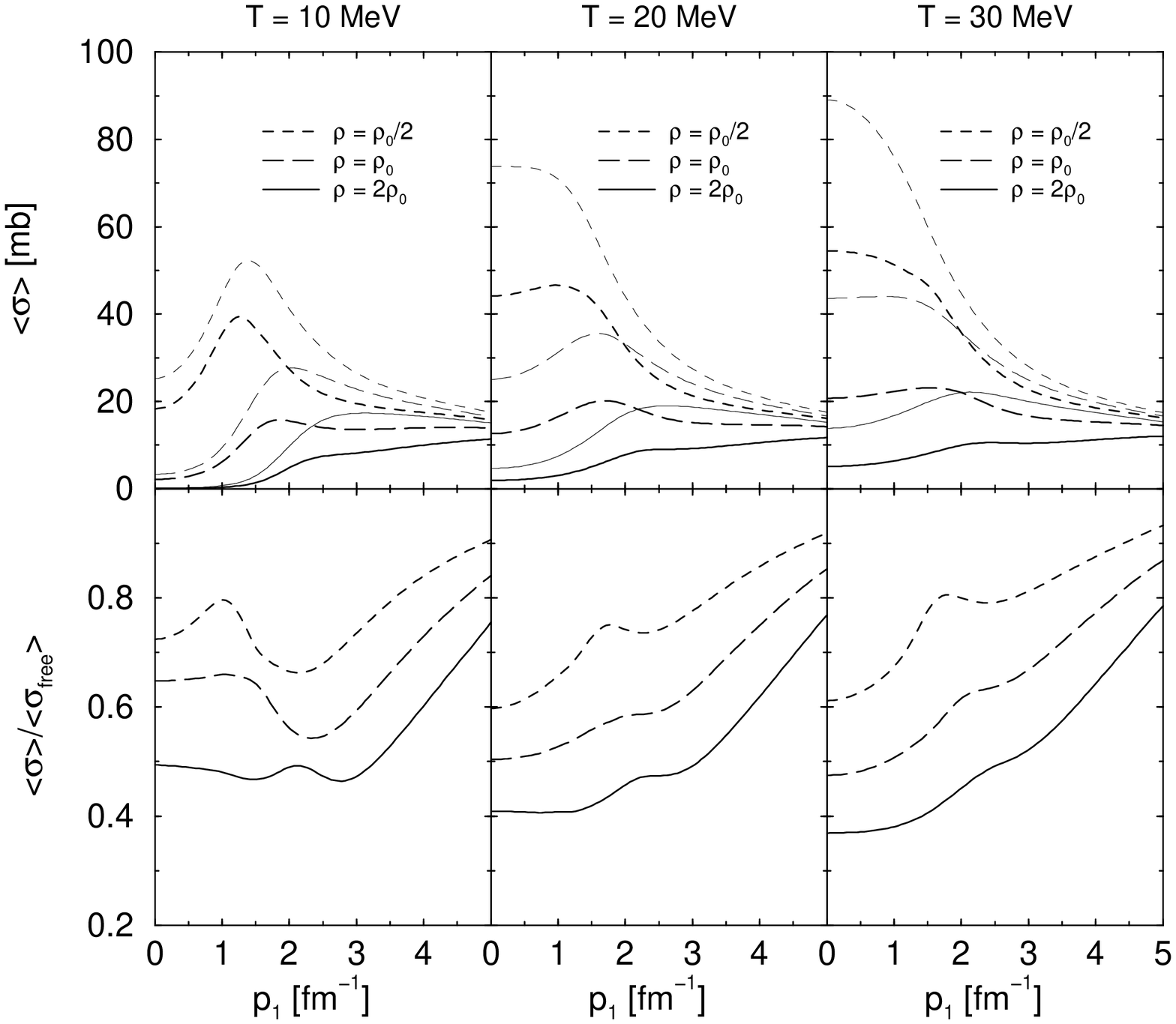,height=7cm}
\caption{Average cross section as function of the momentum $p_1$ for three
values of temperature and density, respectively. Upper half: Absolute values
of the average free (thin lines) and in-medium (thick lines) cross sections.
Lower half: ratio of in-medium vs.\ free average cross section.}
\label{fig-nn_cross}
\end{minipage}
\hfill
\begin{minipage}[t]{8.75cm}
\psfig{figure=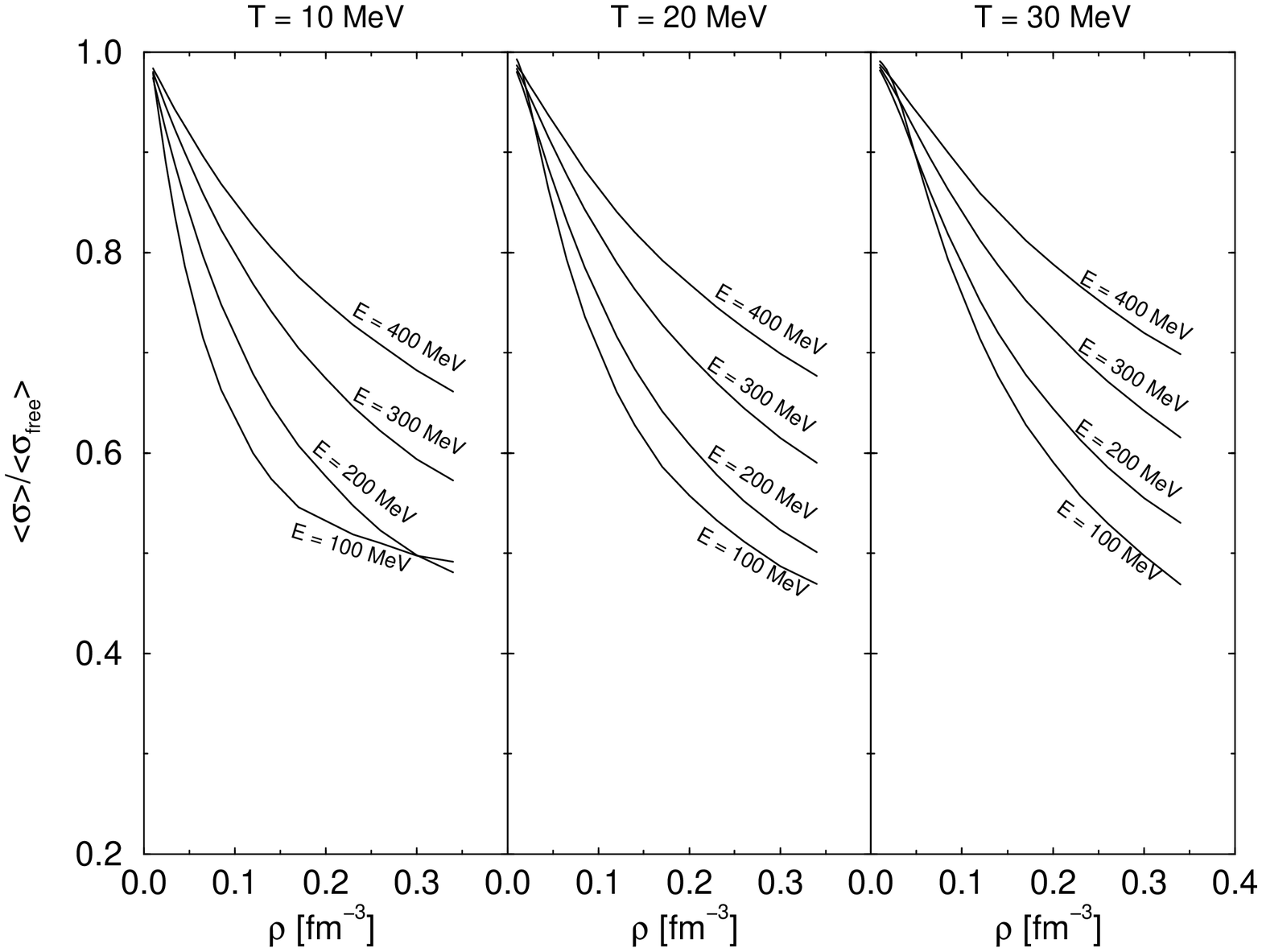,height=6.9cm}
\caption{Ratio of the in-medium vs.\ free average cross section as 
function of the density $\rho$ for three different temperatures and several
values of the scattering energy $E=p_1^2/(2m)$. }
\label{fig-nn_cross_dens}
\end{minipage}
\end{figure}

Another important quantity that can be derived from the two-particle $T$ matrix
is the cross section of elastic nucleon-nucleon scattering. It can be assumed
that the scattering process will be strongly modified in the nuclear medium due to
phase space occupation and mean field effects. This is of particular interest
for the simulation of heavy-ion collisions by transport codes (BUU, QMD) where the
NN cross section is an important input quantity. In terms of the
$T$ matrix it reads
\be
\hspace*{-5mm}
\sigma_T(P,\omega,\mu,T) = \frac{4\pi}{p^2}\frac{N^2(\omega,P)}{(2s_1+1)(2s_2+1)}
\sum_{JLL'}(2J+1)\left|T^{LL'}_{\alpha}(p,P,\omega)\right|^2
\ee
with the generalized density of states $N(\omega,P)$ and the angle-averaged
two-particle energy $\epsilon(p,P)$ (for details see \cite{cross}).
The in-medium total NN cross section is then given by
$\sigma_{NN} = \frac{1}{2}\left[\sigma_{pp}+\sigma_{pn}\right]
=\frac{1}{4}\left[\sigma_0+3\sigma_1\right]$.
In order to reduce the number of variables we use an average procedure according
to collisional integral (loss term) of the Boltzmann equation given by
\bea\label{cross_av}
\langle\sigma\rangle_{\rm\scriptscriptstyle NN}(p_1) & = &
\frac{1}{\langle N\rangle}\int\dn^3p_2 f(p_2)Q_{\rm pp}(p,P)
\frac{\sigma_{\rm\scriptscriptstyle NN}[\epsilon(p,P),P]}{N[\epsilon(p,P),P]}\\
Q_{\rm pp}(p,P) & = & \int\frac{\dn\Omega}{4\pi}[1-f(p_1)][1-f(p_2)]\,,\qquad
\langle N\rangle = \int\dn^3p_2\frac{f(p_2)}{N[\epsilon(p,P),P]}\non\,.
\eea
We obtain an average over the occupation of the final states weighted with the
relative velocity of the particles. Figure \ref{fig-nn_cross} shows in the upper
panels the average cross section as a function of the momentum $p_1$ for various
densities and temperatures. The thin lines represent the calculation of
Eq.~(\ref{cross_av}) using the free cross section instead of the in-medium quantity.
All other quantities in (\ref{cross_av}) remain the same and therefore the 'free'
average cross section $\langle\sigma\rangle_{\rm\scriptscriptstyle free}$ depends
on density and temperature.
It demonstrates
simply the influence of the in-medium cross section compared to the free one,
which can be seen in the lower panels of Fig.~\ref{fig-nn_cross} where the
ratio $\langle\sigma\rangle_{\rm\scriptscriptstyle NN}/
\langle\sigma\rangle_{\rm\scriptscriptstyle free}$ is given. The density dependence
of this ratio is shown in Fig.~\ref{fig-nn_cross_dens} for various scattering
energies and temperatures. The in-medium cross section
is strongly reduced with increasing density, in particular for lower scattering
energies. The temperature dependence is rather weak. The parabolic density
dependence can be easily parametrized as well as the dependence on energy and
temperature (see \cite{cross}).

In heavy-ion collisions not only two-particle states are of relevance, also three
particle processes such as the deuteron formation and break-up play a significant
role. In order to improve the mean-field treatment of the two-particle problem in
nuclear matter, the three-particle problem in matter can be considered.
\begin{figure}[hbt]
\vspace*{-1cm}
\begin{minipage}[t]{8.75cm}
\psfig{figure=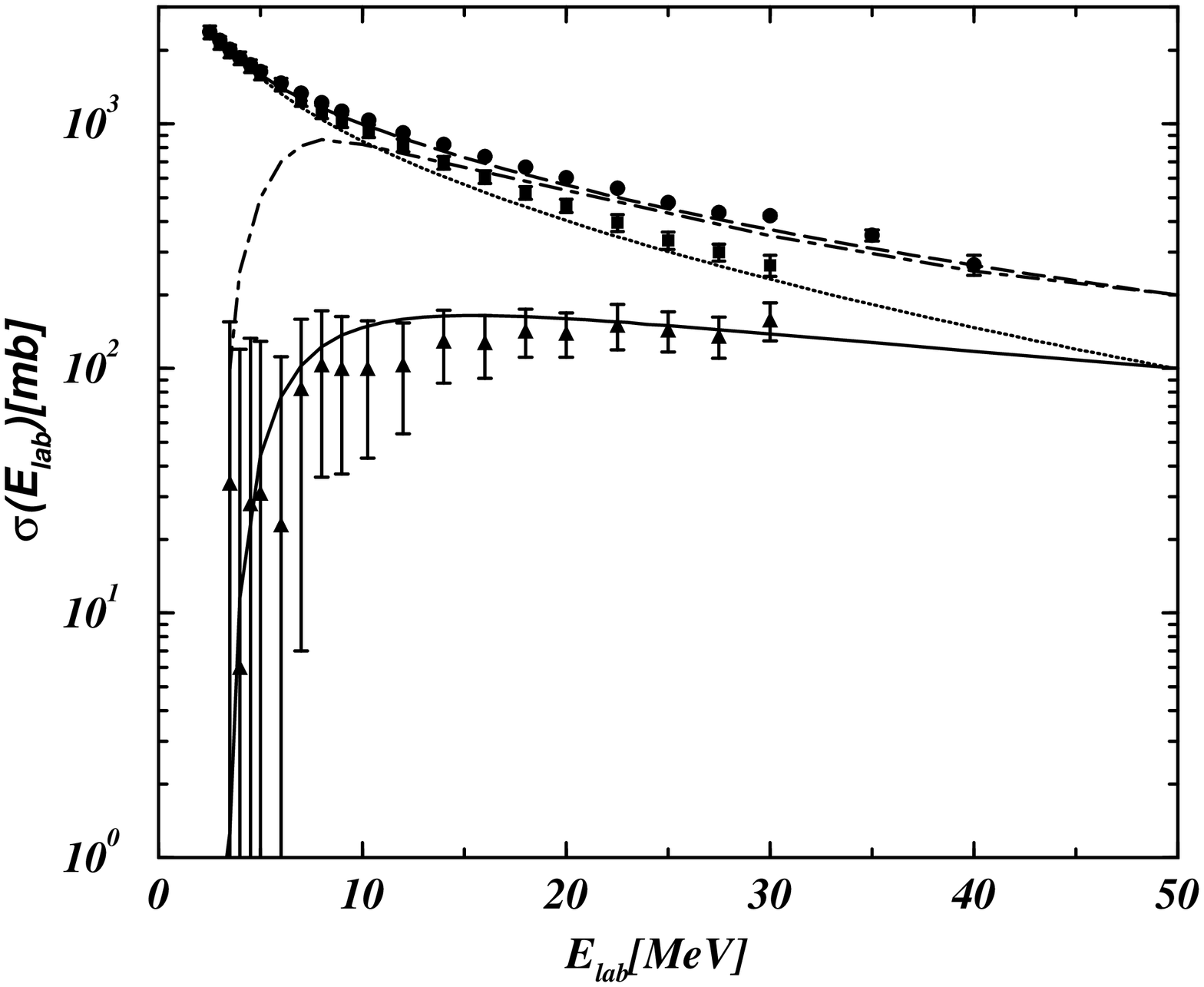,angle=-90,height=7.5cm}
\caption{Neutron-deuteron cross section in nuclear matter (Data from \cite{data}).}
\label{fig-nd_cross}
\end{minipage}
\hfill
\begin{minipage}[t]{8.75cm}
\hspace*{-1cm}
\psfig{figure=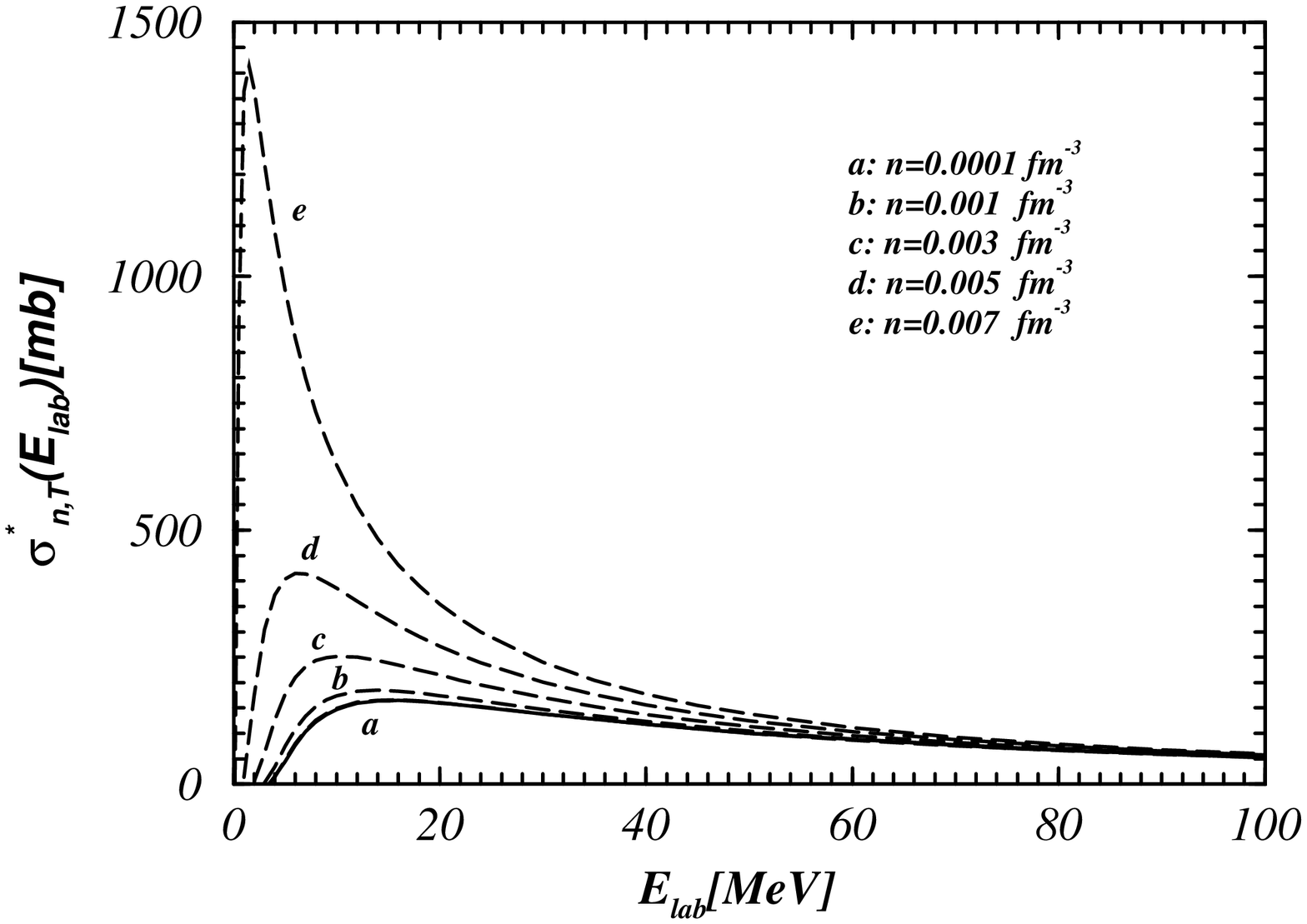,angle=-90,height=8.1cm}
\caption{Medium dependence of the break-up $n-d$ cross section,
temperature $T=10\MeV$ \cite{beyer}.}
\label{fig-med-nd_cross}
\end{minipage}
\end{figure}
In addition to the broadening of the bound state levels, also reaction rates for
break-up and formation of deuterons can be obtained. A Faddeev-type AGS equation
has been solved, where the influence of the medium on the three-particle system
was considered in mean-field approximation \cite{beyer}. As a result an increase
of the deuteron break-up and formation rates has been obtained due to the influence
of the medium.
\section{Superfluidity in nuclear matter}\label{superf}
\begin{figure}[hbt]
\begin{minipage}[t]{8.75cm}
\psfig{figure=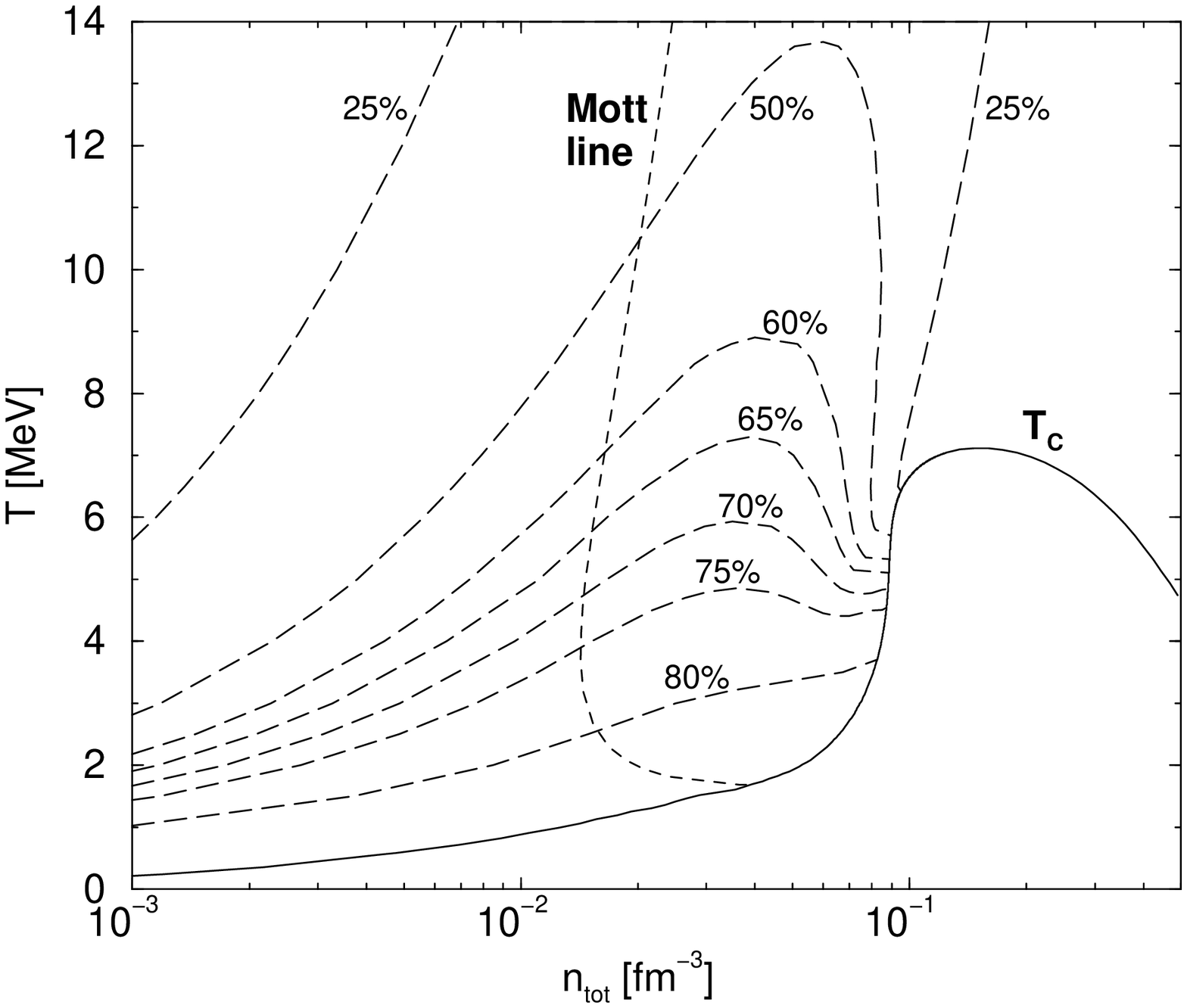,height=7.5cm}
\caption{Composition of nuclear matter in the density-temperature plane
calculated in the generalized Beth-Uhlenbeck approach using the Yamaguchi
potential. The long-dashed curves are lines of equal concentration of
correlated pairs. The Mott line indicates the deuteron break-up due to
Pauli blocking.}
\label{fig-zphys1}
\end{minipage}
\hfill
\begin{minipage}[t]{8.75cm}
\psfig{figure=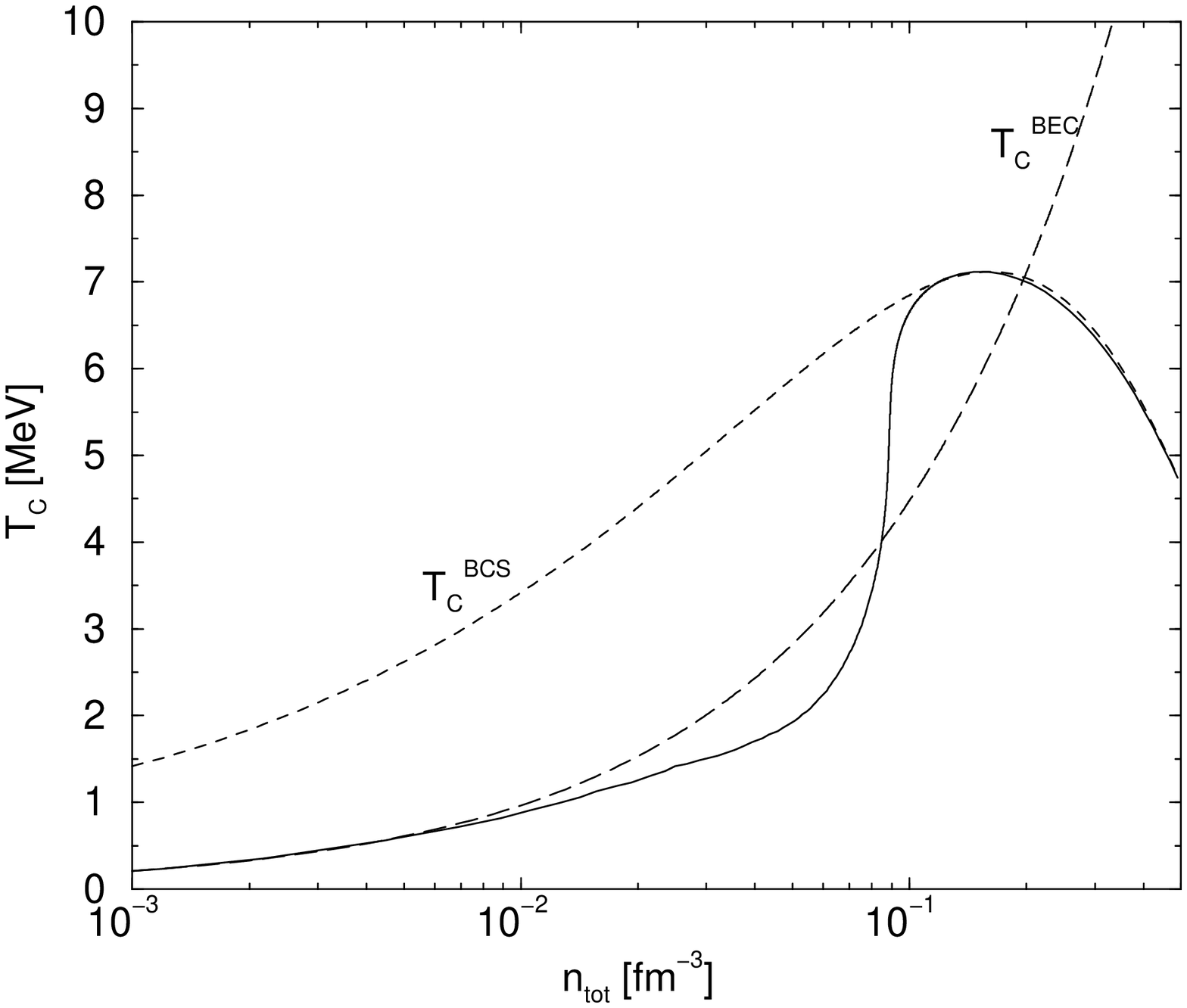,height=7.5cm}
\caption{Critical temperatures of the onset of superfluidity as a function
of the total density. The Beth-Uhlenbeck approximation shows a smooth
transition from Bose-Einstein condensation of deuterons at low densities
(long-dashed curve) to BCS pairing at high densities (short-dashed curve).}
\label{fig-zphys2}
\end{minipage}
\end{figure}

Another highly interesting phenomenon in nuclear matter is the occurrence of
pairing effects. One distinguishes between isospin triplet pairing ($S=0$, $T=1$)
which
is known from even-odd staggering of nuclear binding energies and isospin singlet
pairing ($S=1$, $T=0$) which, however, cannot so easily be identified in nuclear
structure systematics. The reason for this could be the fact that nuclei along
the stability line are isospin asymmetric and therefore protons and neutrons have
different chemical potentials. It has been shown in nuclear matter
calculations \cite{asymm} that asymmetry destroys isospin singlet pairing
very effectively.
In the following we discuss only this kind of pairing in connection with the
formation of a superfluid phase in symmetric nuclear matter.
For this reason we consider the pole of the $T$ matrix describing the Cooper pairs
\be\label{Tc}
\det[1-\re\,J^{\alpha}(\mu,T=T_c,P,\omega=2\mu)]_{ij} = 0\,.
\ee
It defines the critical temperature $T_c$ of the superfluid phase transition which
is a function of the chemical potential (related to the density via the equation
of state) and the total momentum. Using a generalized Beth-Uhlenbeck approach
\cite{SRS90} where the total density is composed of free (quasi-particle) and
correlated (bound and scattering) particles
($n_{\rm tot}=n_{\rm free}+2\,n_{\rm corr}$),
the critical temperature $T_c$ is calculated from (\ref{Tc}) as shown in
Fig.~\ref{fig-zphys1}. Furthermore, the composition ratio
$n_{\rm corr}/n_{\rm tot}$ is given as a function of the total
density and the temperature \cite{zphys}.
In Fig.~\ref{fig-zphys2} $T_c$ is compared to two limiting approximations.
For low densities deuterons can exist (as we have seen previously) and form
a Bose-Einstein condensate below the critical temperature. Treating the
deuterons as ideal Bose system results in the long-dashed curve which is
reached at very low density. At high densities correlation effects are
small (see Fig.~\ref{fig-zphys1}) and we can consider the independent
particle model by introduction of BCS quasi-particles. The result is given
as the short-dashed curve which is reached at high densities. The transition
between both limiting cases is smooth and was first investigated by
Nozi\`{e}res and Schmitt-Rink \cite{NSR}.
\begin{figure}[hbt]
\begin{minipage}[t]{8.5cm}
\psfig{figure=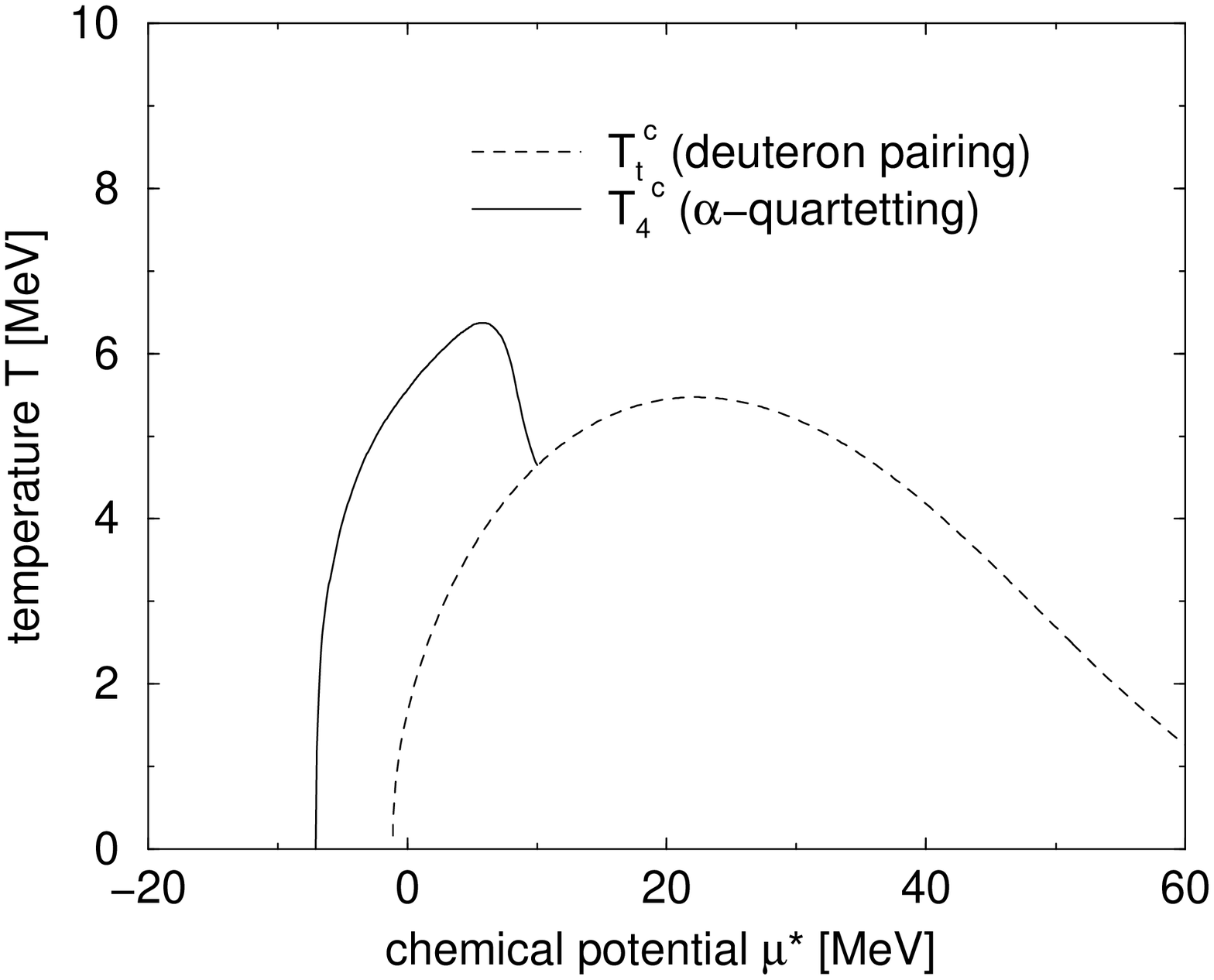,height=7.2cm}
\end{minipage}
\hfill
\begin{minipage}[t]{8.5cm}
\psfig{figure=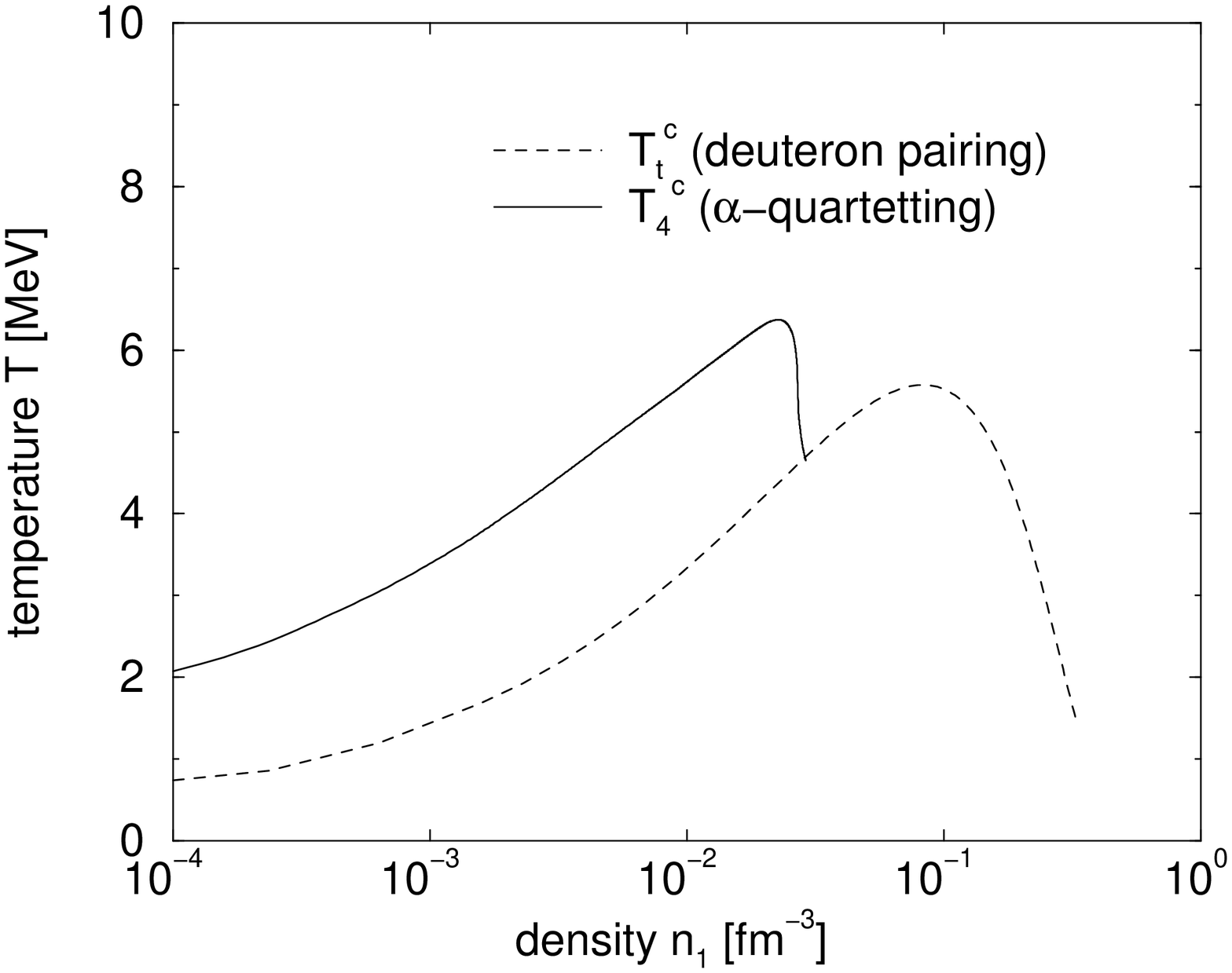,height=7.2cm}
\end{minipage}
\caption{Critical temperatures for the onset of quantum condensation in symmetric
nuclear matter, model calculation \cite{PRL}. The critical temperature of the
onset of two-particle spin triplet pairing $T_t^c$ is compared with $T_4^c$
for the onset of a
four-particle condensate. The left-handed figure shows the behavior as a
function of the chemical potential, the right-handed figure as a function
of the uncorrelated density.}
\label{fig-tc_alpha}
\end{figure}

Besides the two-particle pairing discussed so far, a possible four-particle
condensation (quartetting) may also play an important role in nuclei as well
as during the expansion phase of heavy-ion collisions or in astrophysical
objects. We estimate the onset of quartetting in nuclear matter by the
solution of the following effective four-particle wave equation
\be
\Psi_4(1234) = \sum_{1'2'3'4'}K_4(1234,1'2'3'4',4\mu)\Psi_4(1'2'3'4')\,,
\ee
with the four-particle interaction kernel
\be
K_4(1234,1'2'3'4',z) = V(12,1'2')
\frac{f(1)f(2)}{g_2(12)}\frac{\delta_{33'}\delta_{44'}}{z-E_4(1234)}\,.
\ee
We use the variational calculation ansatz of the wave function $\Psi_4$
\be
\Psi_4(1234) = \phi((\vec{p_1}-\vec{p_2})/2)\phi((\vec{p_3}-\vec{p_4})/2)
\psi(\vec{p_1}+\vec{p_2})
\ee
optimizing the quasi-deuteron wave functions ($\phi$) building the $\alpha$-particle
with a Gaussian ansatz for the relative motion between pairs ($\psi$)
\be
\Psi_4(1234) = \phi((\vec{p_1}-\vec{p_2})/2)\phi((\vec{p_3}-\vec{p_4})/2)
\psi(\vec{p_1}+\vec{p_2})\,.
\ee
The results of this calculation are given in Fig.~\ref{fig-tc_alpha} where the
critical temperatures of quartetting is given as a function of the chemical
potential (left) and the uncorrelated density (right) \cite{PRL}.
Additionally, the pairing curves are given as dashed lines.
\section{Isospin singlet pairing in nuclei}
\begin{figure}[hbt]
\begin{minipage}[t]{8.75cm}
\psfig{figure=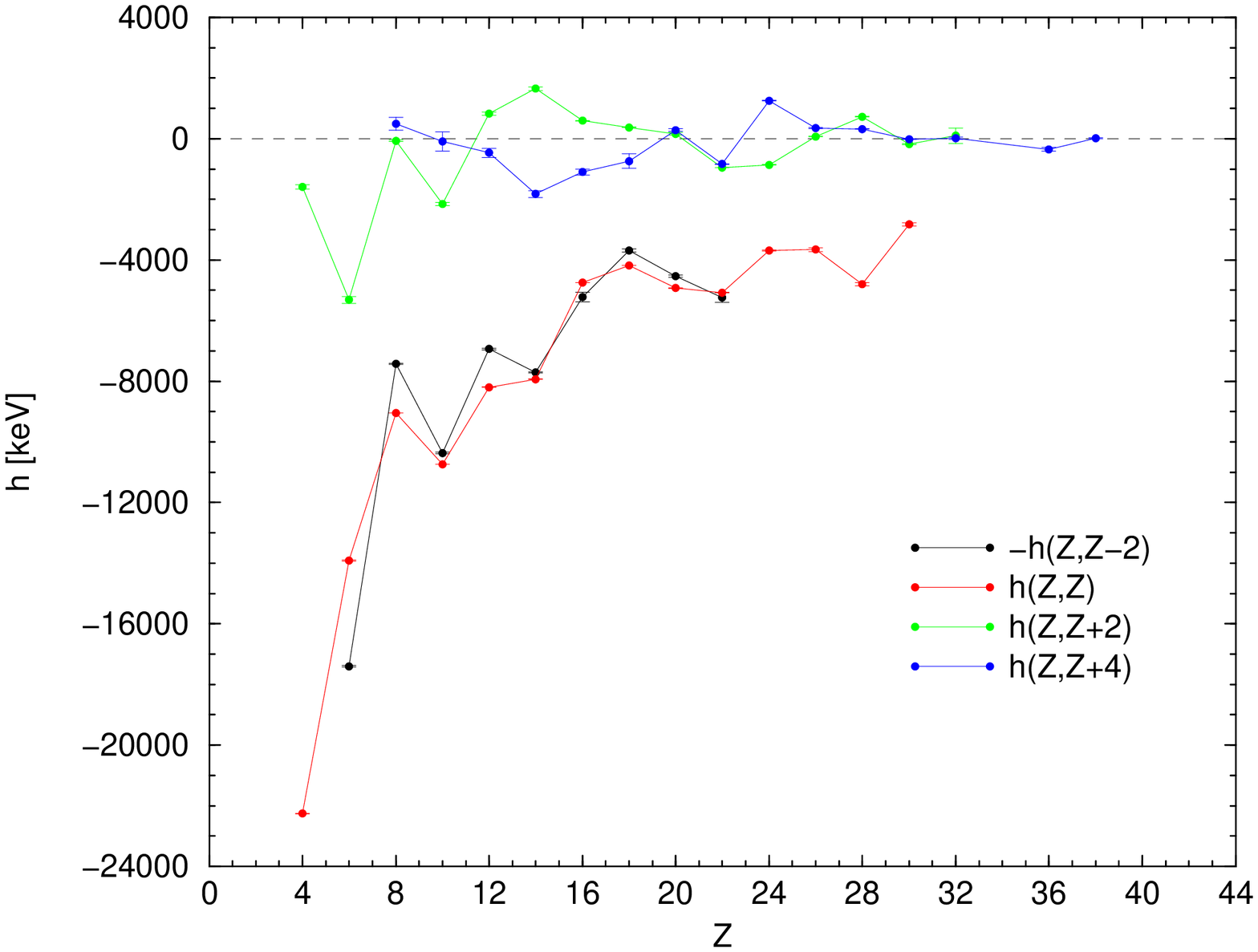,height=6.5cm}
\caption{The filter $h(Z,Z+i)$ for even proton number $Z$ and $i=-2,0,2,4$
considering only even-even nuclei.}
\label{fig-h-g}
\end{minipage}
\hfill
\begin{minipage}[t]{8.75cm}
\psfig{figure=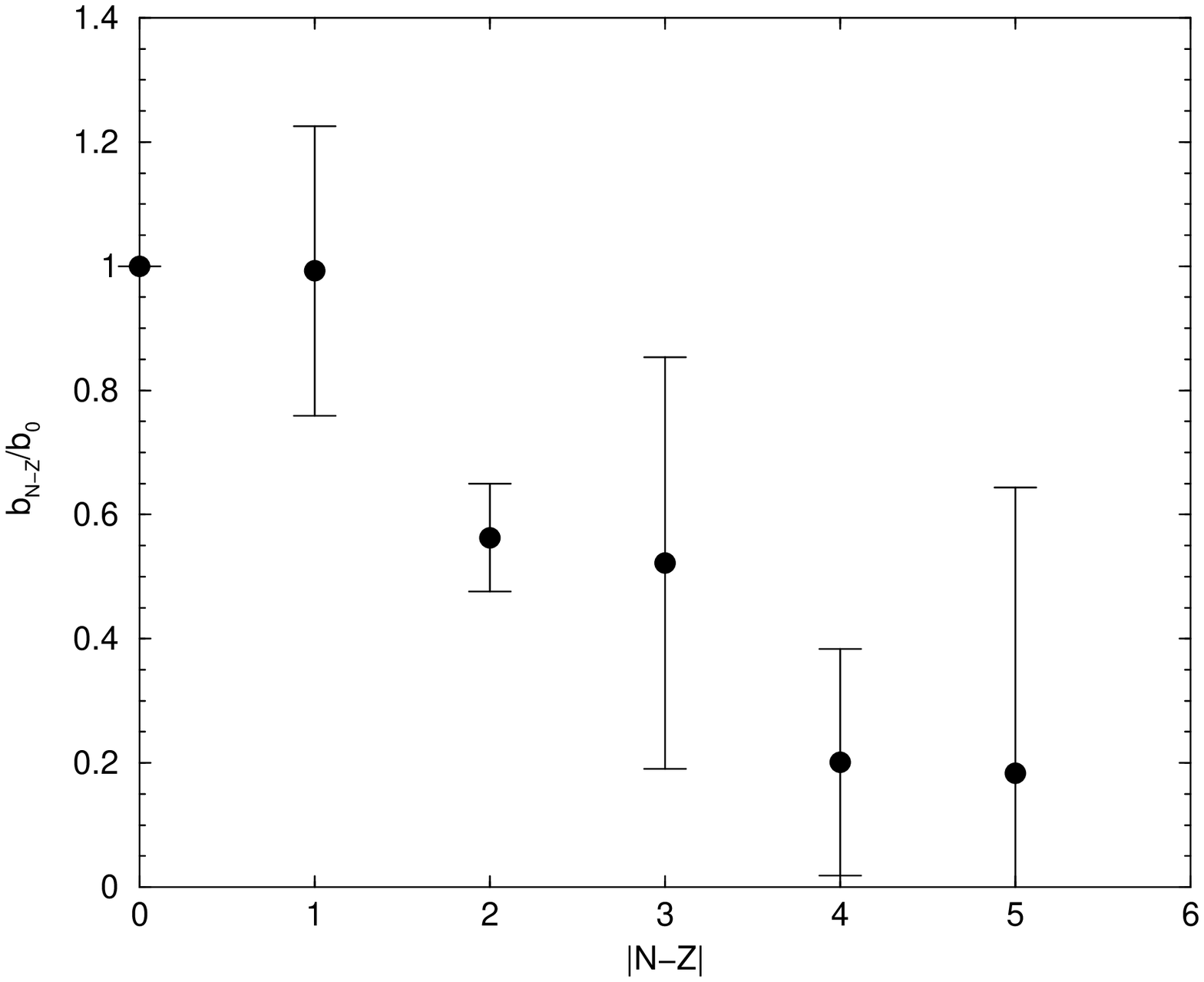,height=6.5cm}
\caption{The average of the ratio $b_{N-Z}/b_0$ as a function of the absolute
value of the proton-neutron difference as it is derived from the filter $h$.}
\label{fig-b}
\end{minipage}
\end{figure}

Finally we come now to the question whether it is possible to find signatures of
isospin singlet pairing and quartetting in the systematics of nuclear
binding energies.
Empirically, the binding energy of nuclei can be decomposed according to the
Bethe-Weizs\"acker formula
\bea
B(Z,N) & = & B_{\rm bulk}(Z,N) + B_{\rm surf}(Z,N) + B_{\rm Coul}(Z,N)
+ B_{\rm asym}(Z,N)\non\\
&&+{}B_{\rm shell}(Z,N) + B_{\rm pair}(Z,N)+\Delta B(Z,N)\label{bwf}
\eea
which contains the main contributions including $B_{\rm pair}$ for $T=1$ pairing.
The additional term $\Delta B(Z,N)$ contains among others the effects of
correlations between protons and neutrons, both isospin singlet pairing and
quartetting. We represent $\Delta B(Z,N)$ as a function of the asymmetry $N-Z$
\bea
\Delta B(Z,N)&=&b_0(Z)\delta_{Z,N}+b_1(Z)\delta_{Z,N+1}+b_{-1}(Z)\delta_{Z,N-1}
+ b_2(Z)\delta_{Z,N+2} + b_{-2}(Z)\delta_{Z,N-2} +\dots\non\\
&=& \sum_i b_i(Z) \delta_{Z,N+i}\label{deltas}
\eea
Now the question arises how to extract $\Delta B(Z,N)$ from experimental binding
energies. For this purpose we define several types of filters which can be used to
cancel the main contributions of (\ref{bwf}) leaving only the term of interest.
The horizontal and vertical filters
\bea
h(Z,N)&=&2 B(Z,N)-B(Z,N-2)-B(Z,N+2)-2B(Z-2,N)+B(Z-2,N-2)\non\\
&&{}+B(Z-2,N+2)\,,\label{h-filter}\\
v(Z,N)&=&2 B(Z,N)-B(Z-2,N)-B(Z+2,N)-2B(Z,N+2)+B(Z-2,N+2)\non\\
&&{}+B(Z+2,N+2)\label{v-filter}
\eea
are differences of second order and the smallest filters possible.
Other filters can be constructed using $h$ and $v$ such as
\be
W(A) = -\frac{1}{8} v(\frac{A}{2}, \frac{A}{2}-2) -\frac{1}{8} h(\frac{A}{2},
\frac{A}{2}-2)\,,
\ee
which is the type of filters Satula and Wyss used \cite{Satula}, or
\be
g(Z,N) = \frac{1}{8} h(Z, N) -\frac{1}{8} h(Z+2, N)\,,
\ee
which is the most symmetric filter. We will concentrate here only on the
$h$-filter. In Fig.~\ref{fig-h-g} $h(Z,Z+i)$ is given for even-even nuclei.
From this and from odd-odd and odd-even/even-odd nuclei, which are not displayed
here, one can derive the following features:
(i) $h(Z,Z)\approx -h(Z,Z-2)$, (ii) $h(Z,Z-1) \approx 0$, and
(iii) $h(Z,Z+i) \approx 0$ for $i>3$.
Based on these properties we can make assumptions which are necessary to find a
solution of the following equation
\bea
h(Z,N)&=&-{}b_{N-Z-2}(Z)+2b_{N-Z}(Z)+b_{N-Z}(Z-2) -b_{N-Z+2}(Z)\non\\
&&{} -2b_{N-Z+2}(Z-2) +b_{N-Z+4}(Z-2)
\eea
which relates the filter $h$ to the coefficients $b_{N-Z}$ of Eq.~(\ref{deltas}).
Using $b_i(Z-2)\approx b_i(Z)$ results in
\be\label{hb}
h(Z,N)\approx -b_{N-Z-2}(Z)+3b_{N-Z}(Z) - 3b_{N-Z+2}(Z) +b_{N-Z+4}(Z)\,.
\ee
$b_i(Z)\approx b_{|i|}(Z)$ reproduces property (i) and (ii), and
$b_i\equiv 0$ for $|i| \geq 6$ is a consequence of property (iii).
By means of
the results of $h(Z,Z+i)$ ($i=0,1\dots 5$) one can solve the system of equations
(\ref{hb}) and obtains $b_i(Z)$. The average of $b_i(Z)/b_0(Z)$ over the proton number
$Z$ is displayed in Fig.~\ref{fig-b} in dependence of the asymmetry parameter
$|N-Z|$. With increasing asymmetry the parameter $b_{N-Z}$ drops to zero rather
quickly. The contribution of proton-neutron or $\alpha$
correlations $\Delta$ can therefore
be assumed to become small with increasing isospin asymmetry as it is expected
from nuclear matter calculations. The contributions of isospin singlet pairing
to nuclear binding energies can be estimated within a local density approximation.
Based on results for the gap energy in asymmetric nuclear matter further studies
are in progress. The contribution of quartetting can be estimated to be of the
order of approximately $0.2\MeV$ for symmetric nuclei with $A\approx 100$,
see \cite{PRL}.
\section{Conclusions}
We have shown that nuclear matter is an interesting example of a strongly
coupled fermion system. In the region of subnuclear densities
($\rho_0/100\le\rho\le\rho_0$) and several MeV temperature strong correlations
occur leading to a variety of in-medium effects which cannot be explained in
terms of a quasi-particle concept. The appropriate description can be given by
spectral functions of the nucleon or even higher clusters. Two-particle
properties, such as the binding energy of the deuteron, scattering phase shifts
and the nucleon-nucleon or deuteron break-up cross section are strongly modified
in the nuclear medium. Taking these into account will lead to different
predictions concerning transport properties in nuclear collisions like nuclear
flow or abundancies of mass fragments. The appearance of quantum condensates in
nuclear matter is another exciting topic. The inclusion of correlations leads
to significant precursor effects of the superfluid phase transition above $T_c$.
Great effort has to be done to generalize the $T$ matrix approach to describe
the equation of state also for the region below $T_c$.

\end{document}